**Nonlinear Bias and the Convective Fisher Equation**

Oliver Schönborn, Rashmi C. Desai, and Dietrich Stauffer*
Department of Physics, University of Toronto, Toronto, Ont., Canada M5S 1A7
* Visiting from St.F.X.University, Antigonish, N.S. Canada B2G 2W5, and Institute for Theoretical Physics, Cologne University, 50923 Köln , Germany

Abstract: We combine random walks, growth and decay, and convection, in a Monte Carlo simulation to model 1D interface dynamics with fluctuations. The continuum limit corresponds to the deterministic Fisher equation with convection. We find qualitatively the same type of asymmetry, as well as velocity difference, for interface profiles moving in opposite directions. However a transition apparent in the mean-field (continuum) limit is not found in the Monte Carlo simulation.

The Fisher equation [1] combines growth, decay, and diffusion, and was originally used to model population growth subject to limited resources. More recently, a nonlinear convective term found in several other pattern-forming systems [2,3] was added to this equation [4]. This convection is phenomenologically justified in the case where an external field has a non-zero component parallel to interface motion, and competes with inertial effects. In one dimension, the new equation, which is called FEC (Fisher Equation with Convection), then reads:

$$\frac{\partial u}{\partial t} = \frac{1}{2}\frac{\partial^2 u}{\partial x^2} + u(1-u) - \mu u \frac{\partial u}{\partial x} \qquad (1)$$

where all quantities are dimensionless, $\mu$ is a positive parameter which cannot be scaled out and serves to tune the relative strength of convection, and the density $u$ will be between 0 and 1 (thus the convection is said to be towards the right; this is different from *ad*vection to the right, which is a linear process). The "mean field" equation (1) neglects random fluctuations, which we include via a Monte Carlo simulation. Also, the nonlinear nature of convection makes its modelling via Monte Carlo non-trivial, an interesting problem in itself. Ref. 4 gives more background literature on the Fisher equation and the FEC.

The Monte Carlo steps are taken as follows. We take $u_i$ to be an integer multiple of $1/n$ (e.g. n=4), so that each site $i$ of our one-dimensional chain carries $u_i\, n$ particles. We take into account the diffusion, $\frac{1}{2}\partial^2 u/\partial x^2$, by a random walk. For the growth and saturation, $u(1-u)$, we increase $u_i$ with probability $1-u_i$: each of the $u_i\, n$ particles produces another particle with probability $1-u_i$. Thus "saturation" means that it becomes more difficult to produce new particles as the $u_i = 1$ value is approached, as when resources are limited. A negative probability corresponds to decreasing $u_i$ with probability $u_i - 1$ to the excess over 1 (although we instead set $u_i = 1$ for speed; this does not change the results). The nonlinear convection term, $\mu u \partial u/\partial x$, is modelled by moving each particle at site $i$ to the right with the *density-dependent* probability $\mu u_{i+1}/2$, thus introducing a *non-linear* bias in the random walk. Making the probability depend on the right neighbor rather than on the site itself eliminates artificial bias due to discretisation. If $\mu > 1$ we rescale the time $t$ by a factor $\mu$ and divide each of the above three probabilities by $\mu$. Denoting the change in density between two time steps at cell $i$ by $\Delta u_i$, the master equation can then be written

$$\Delta u_i = \frac{1}{2}(u_{i+1} - 2u_i + u_{i-1}) + u_i(1 - u_i) - \mu u_i \frac{(u_{i+1} - u_{i-1})}{2}. \qquad (2)$$

The first term is recognized to be the discretized diffusion, the second one the growth-saturation, and the last one is the symmetrized version of the discrete convection (this eliminates artificial bias due to discretisation). Note that the effect of convection will be lost if $n$ is only 1: it will in effect be simple linear advection. It will also disappear when each particle in the system becomes surrounded by empty cells, as is the case when there is no growth process (*e.g.* only diffusion and convection, as in Burgers' Equation [3]). Hence, contrary to diffusion, this algorithm will not yield convection if only one particle is in the system or the system has only 2 levels (0 and 1, i.e $n = 1$), hence we choose $n \geq 2$.

One time unit in our one-dimensional simulations (one Monte Carlo step per site) corresponds to one convection, one diffusion, and one growth attempt (sequentially), each handling all particles in the system at once. The diffusion update was done randomly. The convection was tried sequentially right to left, and randomly, without noticeable difference. Growth was done sequentially left to right (although, due to the local nature of growth, one does not expect this to be important). We started with a nonzero density in the center of the chain and stopped the simulation when the boundaries became occupied. Averages were taken over 1000 or 2000 runs. For random numbers we used the built-in 48-bit linear congruential function (LCF) on our Hewlett-Packard workstation, and verified consistency with the GGL 32-bit LCF, and with a shuffler routine [5] to make the cycle of the random number generator effectively infinite. No differences were found.

Fig.1 and Fig.2 respectively show the density profiles for $n = 4$ and $n = 64$, both for $\mu = 4$, at large enough times that the fronts do not change shape. The profile is not symmetric, as expected due to the presence of convection, which breaks left-right symmetry: the interface thickness on the right is smaller than on the left end of the density plateau. However, this asymmetry is less pronounced than in the solution of eq(1) in ref. 4. The long-time velocities with which the left and the right interfaces propagate are plotted in Fig.3. This may be compared with the mean-field predictions [6]: $\sqrt{2}$ (for left front, independent of $\mu$; for right front, for $\mu \leq \sqrt{2}$) and $1/\mu + \mu/2$ (for right front when $\mu \geq \sqrt{2}$). The important qualitative difference is that in the continuum limit the two fronts have the same speed $\sqrt{2}$ for a certain range of $\mu < \sqrt{2}$, while in the discrete Monte Carlo modelling of convection, the two speeds appear different for all non-zero $\mu$. Understanding the reason for this difference will require further investigation. Often mean field theories predict transitions that disappear when fluctuations are included (e.g. one dimensional Ising model).

We found that reducing the probabilities for diffusion, growth, saturation and convection by the same amount restores the strong asymmetry found in ref.4 between the left and the right interface thickness. This is equivalent to decreasing the time step, thus making the time evolution of the system less discontinuous. This also allows $\mu$ to span a greater domain without having to do any time rescalings. The results in Fig.4 were obtained by reducing probabilities by a factor of 1/8 (so $\mu$ could go continuously from 0 to 8) and looking at $\mu = 8$. The discreteness of the time step then seems crucial in determining the amount of asymmetry convection can give to the profile. No qualitative change was observed in the $\mu$-dependence of the asymptotic front velocity, however. Elsewhere [6] we have shown that in the corresponding mean field description, the asymptotic velocity is constant for $\mu \leq 0.5$. In our present Monte Carlo simulations, it is not clear whether the velocity is really constant or just slowly varying for $\mu \leq 0.5$. A more thorough investigation must be done for small $\mu$.

To summarize and conclude, fluctuations were introduced in a phenomenological model involving diffusion, growth and saturation, and non-linear convection. The convection seems to have been properly modeled in that basic qualitative features of the continuum description are observed both in left-right asymmetry of interface velocities as well as interface profile. The fluctuations become less important when a smaller time-scale is used (obtained by reducing the probabilities by some factor larger than 1). A more thorough evaluation of the effect of discreteness, whether improvements can be made to the modelling of convection via Monte Carlo, as well as studies of a Langevin-type equation (eq.1 with noise) and generalization to higher dimensions, are possible venues worth exploring.

We would like to thank Chuck Yeung for helpful discussions. This work is supported by NSERC of Canada, the Fonds Canadien pour l'Aide à la Recherche, and the Canada Council.

Fig.1: Density profiles for chains of length 500, with $\mu = 4$ and n=4 (average over 1000 realisations).

Fig.2: As Fig.1, for n=64.

Fig.3: Asymptotic velocities of the left and right interface for n=4 and n=64. Rescaled mean field results for eq(1) are shown (right front, solid line; left front, dashed) for *qualitative* comparison, as they give velocity greater or equal to $\sqrt{2}$.

Fig.4: As in Fig.2, but with the probabilities for all four stochastic processes reduced by a factor $1/8$, and $\mu = 8$.

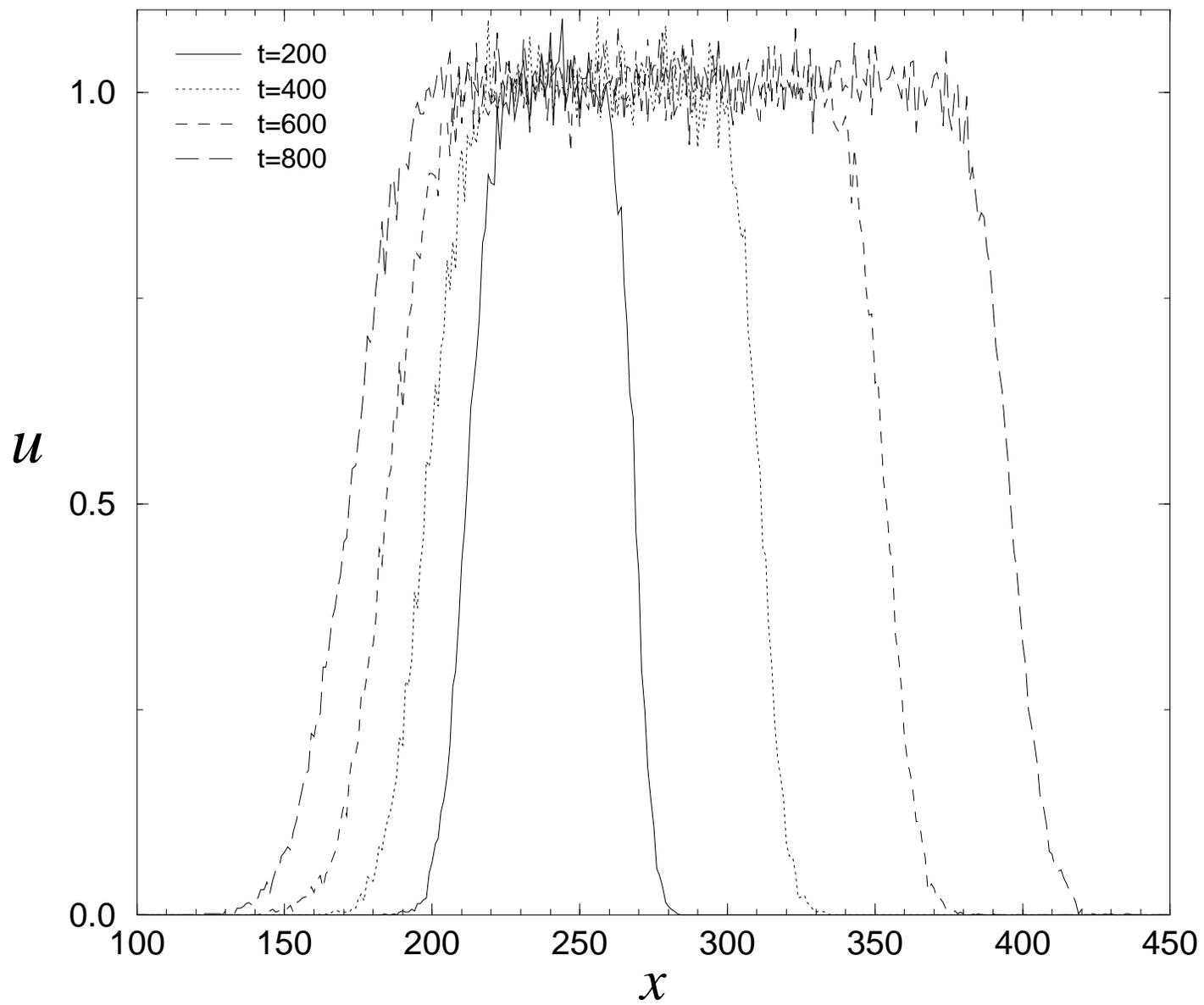

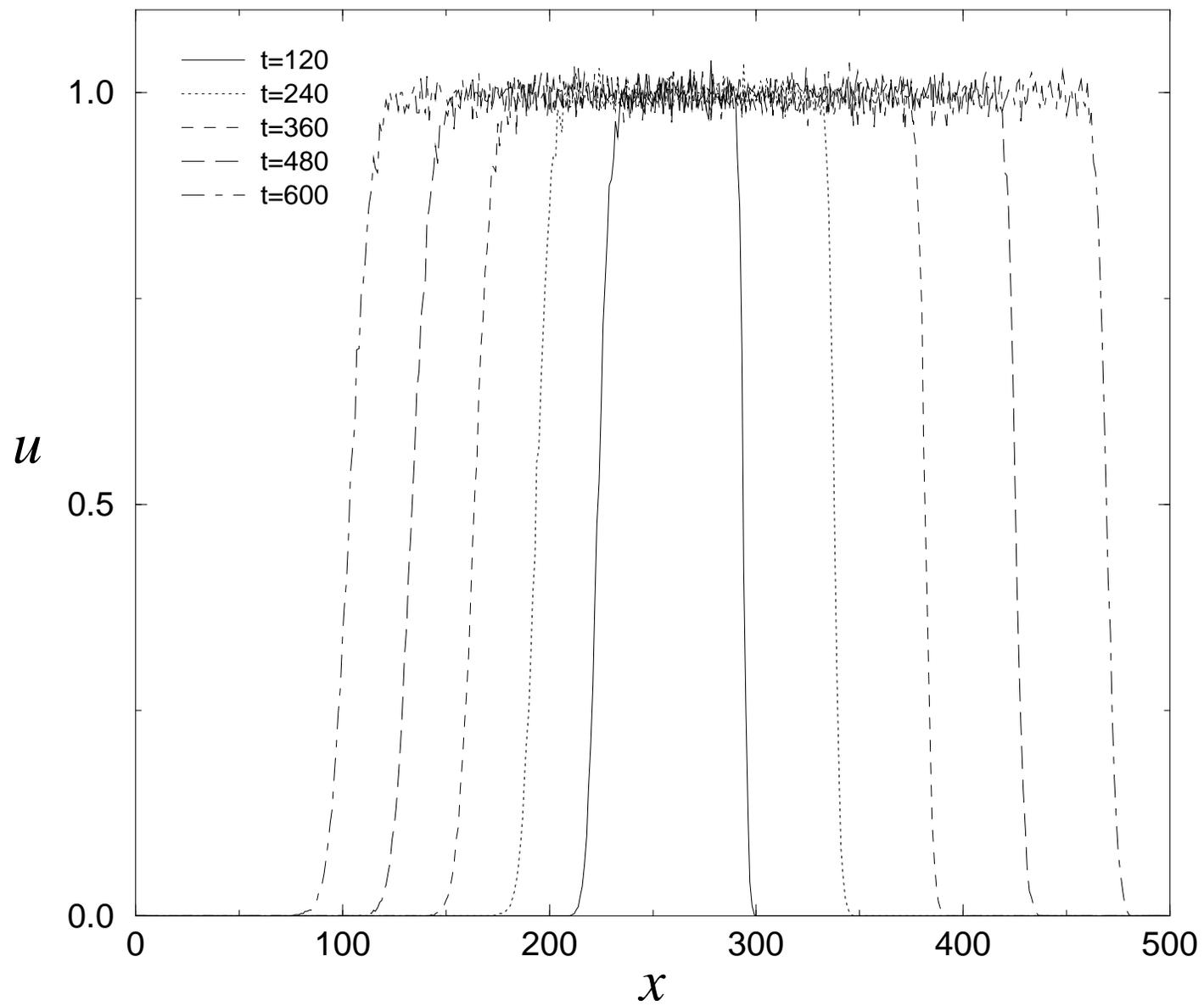

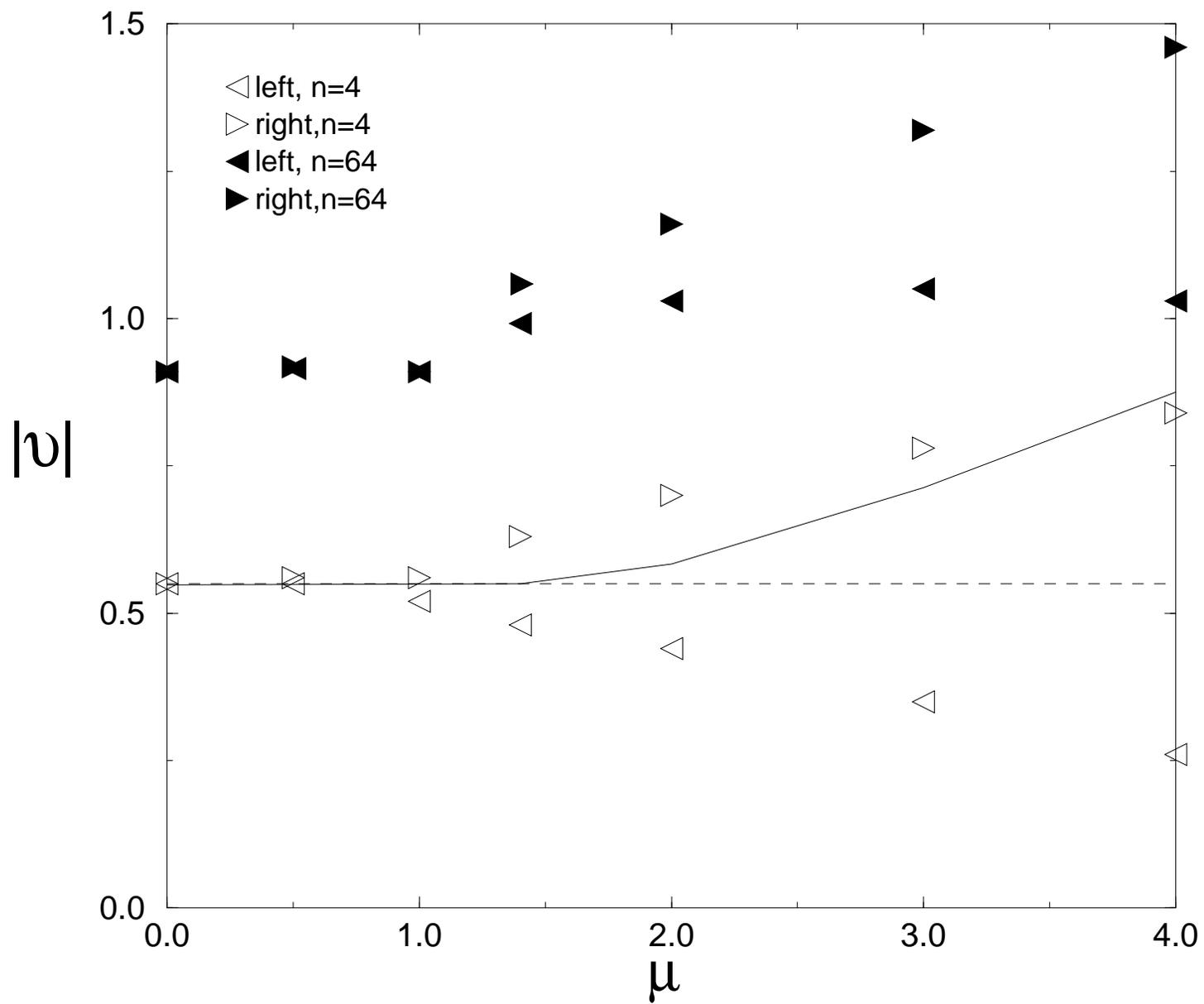

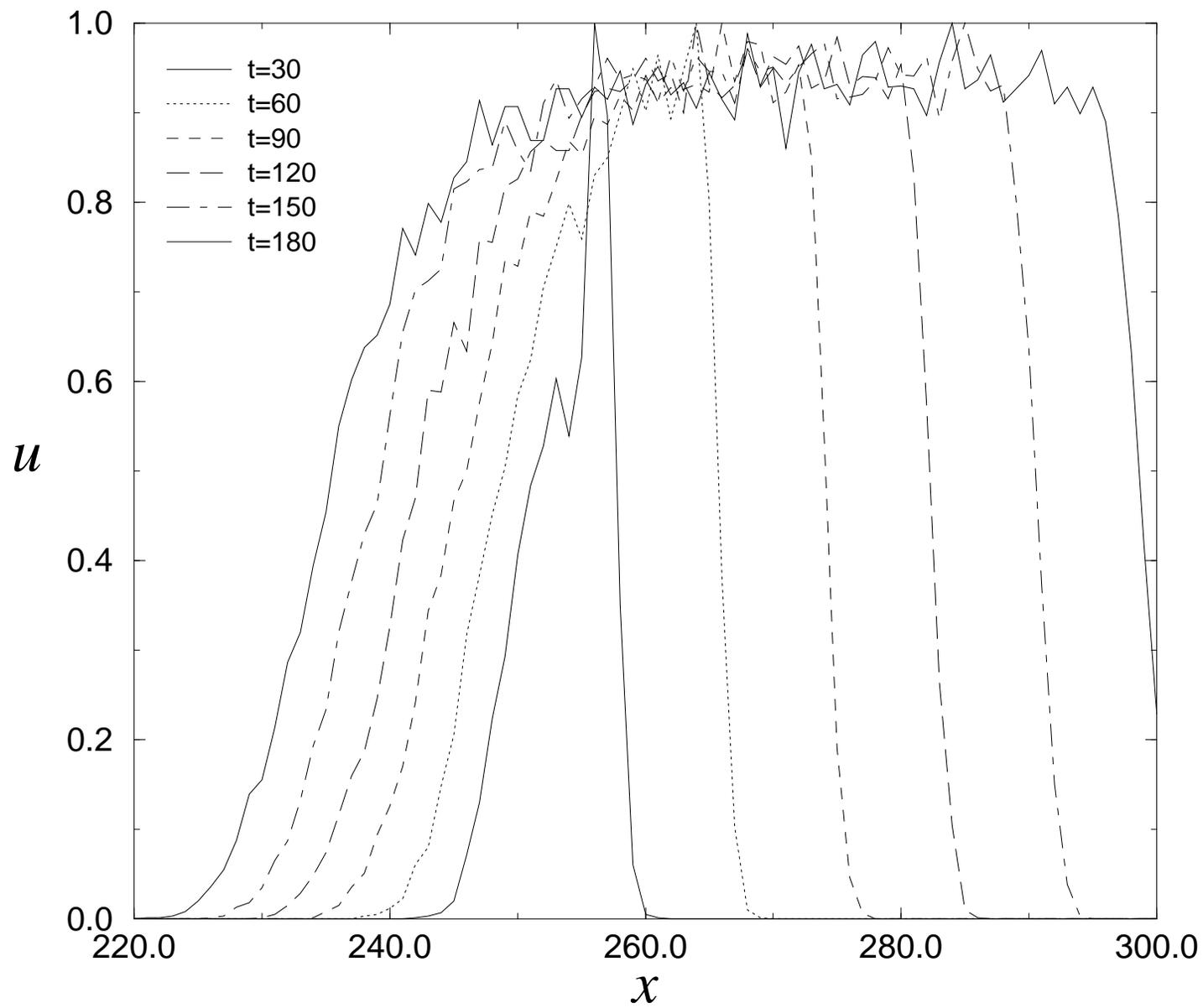